\documentclass[12pt]{iopart}
\usepackage{subcaption}
\expandafter\let\csname equation*\endcsname\relax
\expandafter\let\csname endequation*\endcsname\relax
\usepackage{amsmath}
\usepackage{graphicx}
\usepackage{wasysym}

\begin{document}
\article{{\it J.~Phys.~Commun.}}{Quantum particle in a spherical well confined by a cone}
\author{Raz {Halifa Levi},$^1$ and Yacov Kantor}
\address{Raymond and Beverly Sackler School of Physics and
Astronomy, Tel Aviv University, Tel Aviv 69978, Israel}
\ead{$^1$razhalifa@gmail.com}
\date{\today}

\begin{abstract}
We consider the quantum problem of a particle in either a spherical box or a
finite spherical well confined by a circular cone with an apex angle $2\theta_0$
emanating from the center of the sphere, with $0<\theta_0<\pi$. This non-central
potential can be solved by an extension of techniques used in
spherically-symmetric problems. The angular parts of the eigenstates depend on
azimuthal angle $\varphi$ and polar angle $\theta$ as $P_\lambda^m(\cos\theta){\rm e}^{im\varphi}$
where $P_\lambda^m$ is the associated Legendre function of integer order $m$
and (usually noninteger) degree $\lambda$. There is an infinite discrete set of
values $\lambda=\lambda_i^m$ ($i=0,1,3,\dots$) that depend on $m$ and
$\theta_0$.  Each $\lambda_i^m$ has an infinite sequence of eigenenergies
$E_n(\lambda_i^m)$, with corresponding radial parts of eigenfunctions. In a
spherical box the discrete energy spectrum is determined by the zeros of the
spherical Bessel functions.  For several $\theta_0$ we demonstrate the validity
of Weyl's continuous estimate ${\cal N}_W$ for the exact number of states $\cal N$
up to energy $E$,  and evaluate the fluctuations of $\cal N$ around ${\cal N}_W$.
We examine the behavior of bound states in a well of finite depth $U_0$, and
find the critical value $U_c(\theta_0)$ when  all bound states disappear.
The radial part of the zero energy eigenstate outside the well is $1/r^{\lambda+1}$,
which is not square-integrable for $\lambda\le 1/2$. ($0<\lambda\le 1/2$ can
appear for $\theta_0>\theta_c\approx 0.726\pi$ and has no
parallel in spherically-symmetric potentials.) Bound states have spatial
extent $\xi$ which diverges as a (possibly $\lambda$-dependent) power law
as $U_0$ approaches the value where the eigenenergy of that state vanishes.

\end{abstract}

\section{Introduction}\label{sec:Intro}

Closed form solutions of the non-relativistic Schr{\"o}dinger equation for a single
particle are useful for intuitive understanding of quantum mechanics \cite{Schiff68}.
Unfortunately, exact solutions are not very common. Even in one dimension (1D) the
list of ``simple," analytically solvable, potentials is rather short: it
includes the trivial cases of ``particle in a box" or finite-depth square
well potential, harmonic oscillator, and a list of moderate length of additional
potentials~\cite{Flugge99,Infeld51,Cooper01}, or potentials that can be reduced
to such simple potentials
by appropriate transformations (see, e.g., \cite{Derezinski11} and references therein).
In higher dimensions, ``exactly solvable" problems are usually reduced to
a sequence of 1D problems, such as separation of the $d$-dimensional ``particle in
a rectangular box" problem into $d$ 1D problems in Cartesian coordinates, or similar
separation of a $d$-dimensional harmonic oscillator into 1D oscillators.
(In exceptional cases not amenable to variable separation, alternative methods
based on supersymmetry or shape invariance exist \cite{Cannata02,Ioffe10}.)
For {\em central} potentials, such as Coulomb interaction, or ``spherical box" or
finite spherical well, the simplification is achieved by separating the radial
equation from the angular part, while the angular part in $d\ge3$ can also be separated
into several differential equations corresponding to various angles such as the polar
angle $\theta$ and the azimuthal angle $\varphi$ in three dimensions
(3D)~\cite{Derezinski11}.

In this work we consider a particle either confined in a 3D spherical box
or placed in a finite depth spherical well. In both cases the allowed space is also
confined by a rigid cone of apex angle $2\theta_0$ with the apex located at the
center of the sphere. The resulting potential is {\em not} spherically-symmetric, i.e.
non-central, but it can be solved using a slight extension of central potential
methods which would be used in the absence of the confining cone. The angle $\theta_0$
is a dimensionless parameter that can qualitatively modify the solutions of
Schr{\"o}dinger equation and introduce some features the are absent in the central
potential cases.

Besides the pedagogical value of this particular quantum problem as well as
applicability to small quantum systems with similar geometry, it is also related
to several classical problems:
(a) When $i\partial/\partial t$ in the Schr{\"o}dinger equation is replaced by
$\partial/\partial t$, it  resembles a diffusion equation, with quantum
potential $V({\bf r})$  proportional to particle production or absorption rate
at position ${\bf r}$, while a combination of other constants is proportional
to a diffusion constant; it is one of the simpler forms of the Fokker-Planck
equation~\cite{risken}. (b) For long ideal polymers the partition function $Z$
satisfies an equation resembling the Schr{\"o}dinger equation \cite{Gennes69}
with time replaced by imaginary $iN$, where $N$ is the number of monomers, and
the quantum potential $V$ replaced by the potential of the polymer problem
divided by $k_BT$ (cf., Ref.~\cite{HLKK_PRE96}).
(Sequence of the instantaneous monomer positions ${\bf r}(i)$, where $i$ is the
monomer number, can also be viewed as a time sequence ${\bf r}(t)$ of a diffusing
particle, thus mapping the polymer problem onto a diffusion problem.)  In the
polymer problem the usual dependence of the quantum state with energy $E$
on time $\sim{\rm e}^{iEt/\hbar}$ is replaced by the polymer length dependence
$\sim{\rm e}^{-EN}$, and therefore it is dominated by the ground state. The
presence or absence of bound states in the quantum problem corresponds to
the presence or absence of adsorption in the polymer
problem~\cite{HLKK_PRE96,HLKK_PRE98}.

The 3D problems that are not spherically symmetric are usually
not exactly solvable. However, a particular class of non-central
potentials that has the form \cite{Khare94,Kumari18}
\begin{equation}
V(r,\theta,\varphi)=U(r)+\frac{f(\theta)}{r^2}+\frac{g(\varphi)}{r^2\sin^2\theta}
\label{eq:gen_solvable}
\end{equation}
can be separated in a form resembling central potentials. In classical physics,
such a Hamiltonian has three constants of motion \cite{LL_vol1}, while in
Schr{\"o}dinger equation the parts dependent of $\theta$ and $\varphi$ have the form
that is naturally present when the equation is written in spherical coordinates,
and the resulting equation separates into an azimuthal ($\varphi$-dependent) part
that has a possibly non-integer eigenvalue $m$, which appears in the eigenvalue
equation for the polar ($\theta$-dependent) part with possibly non-integer
eigenvalue $\lambda$. Of course, separation of the Schr{\"o}dinger equation into three
one dimensional equations does not by itself make it exactly solvable,
but for a certain collection of potentials it is possible to express the solutions via
known functions and even provide algebraic expressions for the eigenvalues
\cite{Khare94,Kumari18}. For {\em central} potentials the angular parts have
integer eigenvalues $m$ and $\lambda=\ell$, and the angular functions are
spherical harmonics $Y_{\ell m}(\theta,\varphi)$.

In this work we consider a potential without azimuthal dependence ($g(\varphi)=0$),
thus leaving that part of eigenfunction in the standard form (${\rm e}^{im\varphi}$
with integer $m$) familiar from central potentials \cite{Schiff68}. The polar part
of the potential represents the confinement of a particle inside an infinite circular
cone
\begin{equation}
f(\theta)=\begin{cases}
0, & \theta<\theta_0,\\
\infty, & \text{otherwise}.
\end{cases}
\label{eq:cone}
\end{equation}
Such a potential does not introduce additional energy scales, but forces the polar
part of the eigenfunction to vanish for $\theta=\theta_0$. For the particular case of
$\theta_0=\pi/2$ it represents a repulsive plane. By itself, the infinite conical
surface is {\em length scale-free} and represents an interesting case for many
physical problems described by a Laplacian in the presence of a conical boundary, such
as as problems of heat conduction or diffusion near cones \cite{Carslaw59}, or
polymers attached to conical probes \cite{MKK_EPL96,MKK_PRE86,HK_PRE89,AK_PRE91},
or Casimir forces experienced by conical conductors \cite{Maghrebi11}, or
diffraction of electromagnetic \cite{Felsen57,Northover65,Antipov02,Klinkenbusch07}
and acoustic \cite{Elschner18} waves by conical surfaces.

If the apex of the confining cone is placed in the center of a spherical box
or a finite spherical well of radius $a$, then the angle $\theta_0$ controls the
length scale $a\theta_0$ and therefore strongly influences the eigenstates
of the system. However, a change in $\theta_0$ does not modify the
angular part of the Schr{\"o}dinger equation, but only imposes boundary
conditions on that part of the wavefunction.

In Sec.~\ref{sec:InfiniteWell} we demonstrate the variable separation in
Schr{\"o}dinger equation for a particle in a spherical box, and show the
$\theta_0$-dependence of the angular constants and the energy eigenvalues.
We also study the structure of eigenvalue bunches that are created, and
the behavior of the eigenvalues for $\theta_0$ near $\pi$ or 0.
In Sec.~\ref{sec:Weyl} we verify the validity of a continuous function
that estimates the number of states up to a certain energy $E$, and study
the deviations of the exact results from the continuous estimates.
Similar techniques are used in Sec.~\ref{sec:FiniteWell} to study a
particle in a finite spherical well. Special attention is paid
to the presence or absence of bound states. In Sec.~\ref{sec:Exponents}
we examine the properties of zero-energy eigenstates that appear for
special values of the well depth and show that for large $\theta_0$ some
eigenstates are not normalizable. We show that, when the decreasing
well depth approaches the  condition where eigenenergy of a particular
state vanishes, the spatial extent of the eigenfunction diverges with
an exponent that may depend on $\theta_0$. In Sec.~\ref{sec:Discussion} we
compare some our results with analogous properties of spherically-symmetric
potentials, and also discuss the application of our results to polymer adsorption.

\section{Infinite potential well}\label{sec:InfiniteWell}

\begin{figure}[t!]
\center
\includegraphics[width=6 truecm]{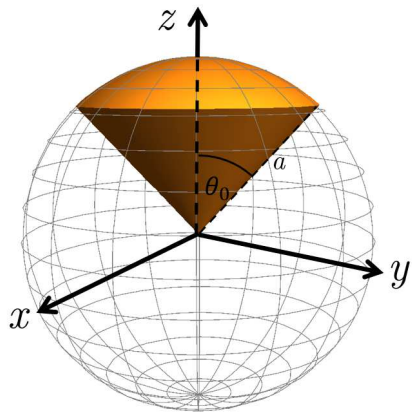}
\caption{Spherical box (infinite potential well) of radius $a$ bounded
by a cone with apex angle $2\theta_0$ (orange volume), with the symmetry
axis along the Cartesian $z$ axis.
}\label{fig:cone_sphere_geometry}
\end{figure}

As a simplest example of a central potential confined by a conical
surface,  we consider a quantum particle of mass $m$ in a 3D spherical
box (infinite potential well) of radius $a$ bounded by a cone with
apex angle $2\theta_0$, such that the complete potential can be written as
\begin{equation}
V\left(\text{\textbf{r}}\right)=\begin{cases}
0, & r<a,\,\theta<\theta_{0},\\
\infty, & \text{otherwise},
\end{cases}
\end{equation}
where the radius $r$ and polar angle $\theta$ are the spherical coordinates.
An example of such confining space is represented in
Fig.~\ref{fig:cone_sphere_geometry}. It is convenient to work with dimensionless
variables, where distances are measured in the units of sphere radius $a$, while the
energies are measured in the units of $\hbar^2 /2ma^2$. The time-independent
Schr{\"o}dinger equation \cite{Schiff68} in these dimensionless variables
is
\begin{equation}
\left(-\nabla^{2}+V\right)\psi=E\psi,
\label{eq:Schrodinger}
\end{equation}
where $E$ is the energy eigenvalue and $\psi$ is the eigenfunction. In the absence
of a confining cone, a spherical box is a textbook example \cite{Schiff68}
of a confined particle. In the presence of a cone, we follow a similar path of solving
the equation  by separation of variables, which in spherical coordinates leads to
\begin{equation}
-\frac{1}{r^2}\left[\frac{\partial}{\partial r}\left(r^2\frac{\partial}{\partial r}\right)
+\frac{1}{\sin\theta}\frac{\partial}{\partial\theta}\left(\sin\theta\frac{\partial}{\partial\theta}\right)
+\frac{1}{\sin^2\theta}\frac{\partial^2}{\partial\varphi^2}\right]\psi\\
+V\psi=E\psi.
\label{eq:Schrod_spherical}
\end{equation}

As in the case of a central potential the solution can be separated
into a product of radial, polar and azimuthal functions
$\psi(r,\theta,\varphi)=R(r)\Theta(\theta)\Phi(\varphi)$. Since, the
potential in Eq.~\eqref{eq:cone} is independent of $\varphi$, the
azimuthal part of the eigenstate satisfies the same equation
as in the case of central potentials
\begin{equation}
-\frac{d^2\Phi}{d\varphi^2}=m^2\Phi,
\label{eq:azimuthal}
\end{equation}
which is satisfied by the functions $\Phi_m(\varphi)={\rm e}^{im\varphi}$ with
$m=0,\pm1,\pm2,\dots$. These are eigenstates of the $z$ component of the angular
momentum since the potential is invariant under rotations around $z$ axis.

The equation for the polar function $\Theta(\theta)$ coincides with the usual
equation used for a central potential, since the restricting cone in
Eq.~\eqref{eq:cone} only influences the boundary conditions ($\Theta(\theta_0)=0$)
but does not otherwise affect the differential equation. The function
$\Theta(\theta)$ obeys the general Legendre equation, in the variable $x=\cos\theta$,
which in terms of polar angle $\theta$ has the form
\begin{equation}
\left[\frac{1}{\sin\theta}\frac{d}{d\theta}\left(\sin\theta\frac{d}{d\theta}\right)
-\frac{m^2}{\sin^{2}\theta}\right]\Theta=-\lambda\left(\lambda+1\right)\Theta,
\label{eq:polar}
\end{equation}
where the eigenvalue $-\lambda(\lambda+1)$ is expressed in terms of the constant
$\lambda$ called the {\em degree} of the equation. In the absence of
a confining cone, the degree of this
equation has integer values $\lambda=\ell$, with $\ell\ge|m|$, and the
eigenfunctions are given by the {\em associated Legendre polynomials} of
$\cos\theta$ of order  $m$ and degree $\ell$, $\Theta(\theta)=P^m_\ell(\cos\theta)$,
and as a result the entire angular part of the eigenfunction is a
spherical harmonic $Y_\ell^m\propto P^m_\ell(\cos\theta){\rm e}^{im\varphi}$
\cite{Schiff68}. While the order $m$ explicitly appears in Eq.~\eqref{eq:polar}, it
only influences the shape of the eigenfunction, but does not affect the integer
degrees $\ell$, and the angular part as well as the energy of the entire
eigenstate remains $(2\ell+1)$-fold degenerate. This is {\em not} the case in the
presence of a confining cone: The associated Legendre polynomial solutions of
Eq.~\eqref{eq:polar} are replaced by the {\em associated Legendre functions}
$P^m_\lambda(\cos\theta)$. For each integer $m$ this equation has an infinite
set of (usually non-integer) degrees $\lambda^m_i$ $(i=0,1,2,\dots)$, such that the
polar function vanishes on the boundary ($P^m_{\lambda^m_i}(\cos\theta_0)=0$).

\begin{figure}[t!]
\center
\includegraphics[width=9 truecm]{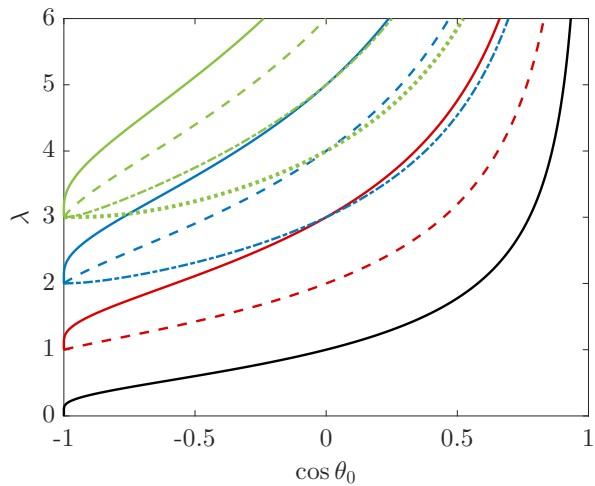}
\caption{Dependence of polar eigenvalues (degrees) $\lambda_i^m$ for various
azimuthal eigenvalues $m$s of a particle in a spherical box, confined by a cone
of apex angle $2\theta_0$, on $w=\cos\theta_0$. Four lowest
line bundles are shown (from bottom to top) $i=0$ (black), $i=1$ (red),
$i=2$ (blue) and $i=3$ (green).  The leftmost point of each bundle corresponds
to the integer degree $\lambda_i^m=i=\ell$. For $w>-1$ each bundle splits into $i=\ell$
separate lines for $m=0,1,\dots,i=\ell$ with (top to bottom in each bundle) $m=0$
depicted as  solid lines, $m=1$ -- dashed lines, $m=2$ -- dot-dashed lines,
and $m=3$ -- dotted line.
When the cone becomes a plane ($w=0$) then the lowest line $\lambda_0^0$
corresponds to integer degree and order $(1,0)$, the second lowest line is
$\lambda_1^1$ at $w=0$ geometry corresponds to $(2,1)$, while
the third lowest point (intersection of $\lambda_1^0$ and $\lambda_2^2$ lines)
corresponds to $(3,0)$ and $(3,2)$.  The order $m$ remains constant along
each of the lines, while the degree $\lambda$ changes.
}
\label{fig:lambda_vs_cos}
\end{figure}

As the angle of the cone changes, and the value of $w=\cos\theta_0$ varies between
 -1 and 1, the geometry of confinement changes between almost unconfined well with
an excluded ``needle" along the negative $z$ axis for $w=-1$, to an excluded cone along the
negative $z$ axis ($-1<w<0$), to confinement by a plane, i.e., particle confined in
the $z>0$ hemisphere ($w=0$), and to a particle confined inside a cone along the
positive $z$ axis ($0<w<1$). If $w$ is changed continuously, the degree $\lambda^m_i$
also changes continuously as depicted in Fig.~\ref{fig:lambda_vs_cos}. For $w=-1$ we
essentially have an unconstrained particle in a spherical box and
$\lambda^m_i=i$ for $i=0, 1, 2,\dots$ independently of the values of $m$, as long
as $i\ge|m|$. These are the $\ell$ values of the spherical harmonics, and each
$\lambda^m_{i=\ell}=\ell$ is degenerate $2\ell+1$ times. This degeneracy is lifted
 once $w$ becomes larger than $-1$, except for two-fold degeneracy for $m\ne0$
for $+m$ and $-m$ pairs, since  $m$ appears only as $m^2$ in Eqs.~\eqref{eq:azimuthal}
and \eqref{eq:polar}. All $\lambda$s monotonically increase with $w$ eventually
diverging in the $w\to 1$ limit.  Thus every value of $\lambda_i(w=-1)$ splits into
$i+1$ branches $\lambda^m_i$ corresponding to different $|m|$s. (We will refer to
each such group of lines as a {\it bundle}.)

The divergence of the $\lambda_i^0$ lines in $w\to1$ limit in
Fig.~\ref{fig:lambda_vs_cos} can be inferred from the properties of the zeros
(roots) $\theta_\lambda^{(j)}$ ($j=1,2,\dots$) of $m=0$ Legendre functions
$P_{\lambda}(\cos\theta)$ in the open interval $(0,\pi)$ of $\theta$s,
i.e. the solutions of $P_\lambda(\cos\theta_\lambda^{(j)})=0$. (For
$\ell<\lambda\le\ell+1$ there are $\ell+1$ such roots.) In fact, all
the curves in Fig.~\ref{fig:lambda_vs_cos} have been constructed by
choosing fixed $\lambda$ and fixed $m$ an finding all the roots, when every
root belongs to a different curve in the figure, and then tracing the
curves by gradually varying $\lambda$. For integer
$\lambda$s only a discrete set of $\theta_\lambda^{(j)}$ can be
accommodated, but for general $\lambda$s the roots can have any value.
This statement can also be inverted to say that any value of
$\theta_0$ can be a root corresponding to an infinite sequence
of $\lambda_i^0$s or any vertical line in Fig.~\ref{fig:lambda_vs_cos}
intersects infinity of $\lambda_i^0$ curves. Several tight bounds on
the roots are known -- see, e.g., Ref.~\cite{Szego36} and references
therein. They can be used to produce large-$\lambda$ approximation
$\lambda_i^0\approx \pi a_i/\theta_0$, with $i+1/2<a_i<i+1$, where the
bounds  on $a_i$ are derived from the bounds on the position of $(i+1)$th
root. (Strictly speaking, the bounds on the roots in \cite{Szego36}
have been derived for integer $\lambda$s, but for small $\theta_0$
and large $\lambda$ they can be used for noninteger $\lambda$s.)
Since for small $\theta_0$, we can approximate $\theta_0\approx\sqrt{2(1-w)}$,
and the functional dependence of the branches becomes
$\lambda_i^0\approx\pi a_i/\sqrt{2(1-w)}$, thus explaining the divergences seen in
Fig.~\ref{fig:lambda_vs_cos}. The bounds on the coefficients $a_i$ also ensure
that the different branches $\lambda_i^0$ do not intersect. (Non-intersection of
$\lambda_i^0$ lines is also ensured by the fact, that an intersection would create
a multiple root of $P_\lambda$ thus contradicting the known fact that all their
roots are simple.)

From the definition of associated Legendre functions $P_\lambda^m$
via derivatives of regular ($m=0$) Legendre functions $P_\lambda$, or from
standard recurrence relations between the functions \cite{HigherTrFuncVol1_Bateman53},
one finds that
\begin{equation}
P_\lambda^{m+1}=-\sqrt{1-x^2}\frac{dP_\lambda^m}{dx}+\frac{mx}{\sqrt{1-x^2}}P_\lambda^m\ .
\label{eq:recur}
\end{equation}
At two consecutive zeros (simple roots) of $P_\lambda^m$ the second  terms
on the right hand side of  Eq.~\eqref{eq:recur} vanish, while the first terms
(the derivatives) have opposite signs, and therefore $P_\lambda^{m+1}$
will have opposite signs at those points. Thus, the zeros of $P_\lambda^{m+1}$
lie in between the zeros of $P_\lambda^m$. Consequently, the branch of
$\lambda_i^m$ in Fig.~\ref{fig:lambda_vs_cos} will be locked between branches
$\lambda_i^{m-1}$ and $\lambda_{i-1}^{m-1}$, if they both exist, and will not
intersect  with them. Thus $m=1$ branches will be between $m=0$ branches, and
$m=2$ branches will be between $m=1$ branches, etc. (However, $m=2$ branch
{\it can} intersect $m=0$ branch, as can be seen in the intersection of
$\lambda_3^2$ and $\lambda_2^0$ lines in Fig.~\ref{fig:lambda_vs_cos}.)
Nevertheless, it means that lines $\lambda_i^m$ with {\em any} $m$ diverge
as $1/\sqrt{1-w}$ for $w\to1$, i.e. have the same divergence as $m=0$
branches. At every integer level $\lambda=\ell$, the
horizontal line in Fig.~\ref{fig:lambda_vs_cos} will cut $\ell-m$ branches
with that particular $m$, since this is the  number of zeros of $P_\ell^m$
in the open interval $(-1,1)$ of $w$. (This excludes  extra zero at $w=-1$).
For a noninteger $\lambda$ between some $\ell$ and $\ell+1$,  the number of
such intersections is $\ell-m+1$.

The eigenfunctions must vanish on the cone boundaries, even in the limit where the
excluding cone becomes needle-like along the negative $z$ axis and, eventually,
just a line for $w\to -1$. For $m\ne0$ the polar eigenfunction of an unrestricted
sphere $P_\ell^m$ vanishes at $\theta=\pi$ and therefore $P_\lambda^m$ naturally
approaches $P_\ell^m$ as $\theta_0$ approaches $\pi$. Consequently, all $m\ne0$
curves in Fig.~\ref{fig:lambda_vs_cos} approach $w=-1$ points linearly. This is not
the case for $m=0$, where $P_\ell^0$ {\em does not vanish} at $\theta=\pi$, and
differs from $P^0_\lambda(\cos\theta_0)=0$ for $w=\cos\theta_0=-1+\epsilon$, with
$0<\epsilon\ll1$. As $w$ approaches $-1$ the restricted and unrestricted solutions
become almost identical everywhere except for a very narrow region around the
negative $z$ axis, that remains present although its volume vanishes. This
behavior is reflected in the fact that the $\lambda_{i=\ell}$ approaches its
limiting value $\ell$ almost vertically: From the asymptotic forms of $P_\lambda^0$
near the singularity \cite{NISTlib} one finds  that for $w$ close to $-1$ the eigenvalue
$\lambda^0_{i=\ell}\approx\ell-1/\ln(1+w)$.

To gain some intuition into the behavior of the curves in Fig.~\ref{fig:lambda_vs_cos}
we examine the relations between the eigenstates of unconfined particles in a spherical
box ($w=-1$) and particles confined in a hemisphere ($w=0$). We note, that $\lambda$s
are integers and with significant degeneracy for $w=-1$, but also for $w=0$ the
$\lambda$s are integers, and there is some degeneracy due to intersections of
different branches. In the former case the polar eigenvalues $\ell=0,1,\dots$, are
degenerate $2\ell+1$ times, since for each $\ell$ we have $m=0,\pm 1,\dots,\pm\ell$.
Spherical harmonics  $Y_{\ell}^m$ can also be used to build the eigenstates of a
hemisphere. We note that  equality $P_\ell^m(\cos\theta_0=0)=0$ is valid when $\ell$
and $m$ have opposite parity. Thus almost half of the eigenfunctions of an
unrestricted sphere vanish on the plane $\theta_0=\pi/2$ (or $z=0$) and can be
used as a set of angular functions for hemisphere. Thus we have eigenfunctions and
eigenvalues with $(\ell,m)=(1,0),\ (2,\pm1),\ (3,0),\ (3,\pm 2),\dots$ The seeming
reduction in the number of the eigenstates reflects the decrease in the volume
of the system. We can now observe how the lowest branch that begins with
$\lambda^0_0(w=-1)=0$, which corresponds to $(\ell,m)=(0,0)$ state of the complete
sphere, increases with increasing $w$ and reaches $\lambda^0_0(w=0)=1$, which
corresponds to $(\ell,m)=(1,0)$ eigenstate of the hemisphere. Similarly, the
$m=1$ branch of $\lambda^1_1(w=-1)=1$ which corresponds to  $(\ell,m)=(1,1)$
of a complete sphere increases and reaches value $\lambda^1_1(w=0)=2$,
which corresponds to $(\ell,m)=(2,1)$ eigenstate of the hemisphere. The $m=0$
branch that also begins at $\lambda^0_1(w=-1)=1$, which corresponds to
$(\ell,m)=(1,0)$ of a complete sphere reaches value $\lambda^0_1(w=0)=3$,
which corresponds to $(\ell,m)=(3,0)$ eigenstate of the hemisphere. At $w=0$ the
latter branch intersects $m=2$ branch that started at $\lambda^2_2(w=-1)=2$ and
reached $\lambda^2_2(w=0)=3$, which corresponds to $(\ell,m)=(3,2)$ eigenstate of
the hemisphere, and therefore completes the eigenstate mentioned before with $m=0$.
Thus, increase in $w$ causes ``reordering" of the eigenstates. There are additional
intersections of different branches, corresponding to a variety of cone angles
$\theta_0$. Those, however, are accidental degeneracies of unrelated states.

The line intersections in Fig.~\ref{fig:lambda_vs_cos} described in the previous
paragraph are in line with the theorem \cite{Lacroix84} that two associated
Legendre functions $P_\ell^m(w)$ and $P_\ell^{m'}(w)$ with integer degrees and orders
and $|m|\ne|m'|$ have no common zeros with exception of the case when $\ell$ and
$m$ have opposite parity, as well as $\ell$ and $m'$ have opposite parity, in
which case they have common zeros at $w=0$. This exactly describes the intersections
at $w=0$ in Fig.~\ref{fig:lambda_vs_cos}. At the same time, it means that there
can be {\em no other intersections at integer $\lambda$s}. Indeed, the accidental
intersection of $\lambda_3^3$ and $\lambda_2^0$ lines in Fig.~\ref{fig:lambda_vs_cos}
appears at non-integer $\lambda$ slightly larger than 3.

\begin{figure}[t!]
\center
\includegraphics[width=9 truecm]{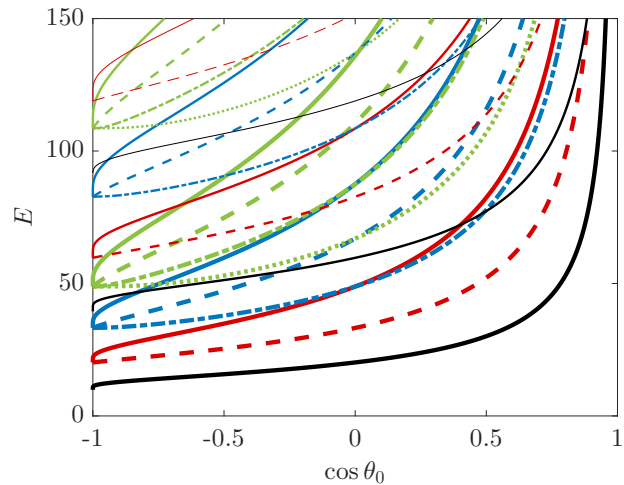}
\caption{The energy $E$ spectrum as function of $w=\cos\theta_0$. For
each $\lambda$ shown in Fig.~\ref{fig:lambda_vs_cos} there is an infinite
sequence of roots $n=1,2,\dots$ shown with different line thicknesses,
the very thick (lowest) line ($n=1$), intermediate thickness (higher)
line ($n=2$), thin (highest) line ($n=3$), while the line styles and
colors are the same is in Fig.~\ref{fig:lambda_vs_cos}. Since the roots
of Bessel functions depend only on the values of $\lambda$ every bundle
of $\lambda$s in Fig.~\ref{fig:lambda_vs_cos} creates multiple similar
bundles of energy lines in this figure. Degeneracies of $\lambda$ lines
in Fig.~\ref{fig:lambda_vs_cos} translate into energy degeneracies in
this figure. Additional accidental energy degeneracies are created by
intersection energy lines belonging to different $n$s.
}
\label{fig:E_vs_cos}
\end{figure}

The radial part $R(r)$ of the eigenfunctions of Eq.~\eqref{eq:Schrod_spherical} is
a solution of the equation
\begin{equation}
\frac{1}{r^2}\frac{d}{dr}\left(r^2\frac{dR}{dr}\right)
+\frac{\lambda(\lambda+1)}{r^2}R=\left[E-U(r)\right]R,
\label{eq:radial}
\end{equation}
where the radial component $U(r)$ of the potential $V({\bf r})$ vanishes inside
the well, and only manifests by the boundary condition $R(r=1)=0$. This is
correct both in the absence and the presence of the confining cone. We note that
the equation only depends on the polar eigenvalue $\lambda$ but not on $m$, although
the actual value of $\lambda$ may depend on $m$. This equation is solved
by the spherical Bessel functions $j_{\lambda}(\sqrt{E}r)$, so that the complete
eigenfunctions are $\psi\left(r,\theta,\varphi\right)\propto j_{\lambda}
\left(\sqrt{E}r\right)P_{\lambda}^{m}\left(\cos\theta\right)e^{im\varphi}$.
By imposing the boundary condition $j_\lambda (\sqrt{E})=0$, we find the energy
spectrum of the system
\begin{equation}
E_n(\lambda)=\alpha_n^2(\lambda),
\label{eq:E_relate_to_root}
\end{equation}
where $\alpha_n(\lambda)$ is the $n$th zero of $j_\lambda(x)$.
Figure~\ref{fig:E_vs_cos} shows part of the energy spectrum as function of
$\cos\theta_0$ for different values of $n$, $\lambda$ and $|m|$. Since the
eigenenergies directly depend only on $\lambda$, the Fig.~\ref{fig:E_vs_cos}
slightly resembles Fig.~\ref{fig:lambda_vs_cos}. In particular
degeneracies (accidental or not) seen in Fig.~\ref{fig:lambda_vs_cos} are
``reproduced" in the energy lines. To facilitate comparison of these two
figures, we  employed the same coloring and line-type scheme for the graphs: the
energy line types and colors are identical to lines types and colors
used to described $\lambda$s for which those energies were calculated.
However, since every specific value of  $\lambda$ produces an infinite series
of roots $\alpha_n(\lambda)$, with $n=1,2\dots$, every single line in
Fig.~\ref{fig:lambda_vs_cos} produces many lines in Fig.~\ref{fig:E_vs_cos}.
They are distinguished in Fig.~\ref{fig:E_vs_cos} by {\em line thickness}
where $n=1$ corresponds to the thickest lines, and the thickness decreases
with increasing $n$. Besides the energy degeneracies originating in the degeneracies
of $\lambda$s, there are additional accidental degeneracies when roots of
different order belonging to different $\lambda$s coincide. Due to simple
relation between the energies and $\lambda$s in Eq.~\eqref{eq:E_relate_to_root}
the properties of $\lambda$s in $w\to-1$ and $w\to1$ limits are
partially (qualitatively) mimicked in the energy spectrum.

Since for large $\lambda$ the first root $\alpha_1(\lambda)\sim\lambda$ \cite{Arriola89},
the ground state energy will (due to Eq.~\eqref{eq:E_relate_to_root}) diverge
in the $w\to1$ limit as  $(\lambda_0^0)^2\sim1/\theta_0^2\sim 1/(1-w)$.
The same conclusion can be reached from Heisenberg's uncertainty principle,
since for small $\theta_0$ the confining dimension in the sphere of
radius $a$ constrained by the cone is $a\theta_0$ leading to uncertainty
in momentum of order $\hbar/a\theta_0$, which corresponds in our
dimensionless units to energy $E\sim1/\theta_0^2$.

\section{Weyl relations in a confined infinite well}\label{sec:Weyl}

As can be seen in Fig.~\ref{fig:E_vs_cos} the eigenenergies corresponding to
a particular branch increase with increasing confinement, and the number of
eigenstates ${\cal N}(E)$ with energies below a certain value $E$ decreases.
The exact ${\cal N}(E)$ in a box of an {\em arbitrary shape} is a step-wise
function, which jumps upwards by an integer amount whenever an eigenenergy
is encountered. The size of the jump is the degeneracy of that energy level.
For more than a
century Weyl and his successors developed a smooth function ${\cal N}_W(E)$
approximating ${\cal N}(E)$, which in $d$ dimensions has the form
${\cal N}_W=a_1E^{d/2}+a_2E^{(d-1)/2}+a_3E^{(d-3)/2}+\dots$.
(For an overview see Refs.~\cite{Baltes76,Arendt09,Ivrii16}.)  The expression for
${\cal N}_W$ is rather general and requires surprisingly little information
about the system. However, some uncertainty exists both regarding the
exact conditions  for the validity of such expressions and the behavior of
the {\em remainder}
\begin{equation}
r(E)\equiv{\cal N}(E)-{\cal N}_W(E).
\label{eq:remainder}
\end{equation}
The subject has been extensively studied for a particle in a square (cubic)
box in 2D (3D), where the problem of  number of states is reduced to the
counting of number of square (cubic) lattice points $\cal N$ within a
circle (sphere) of radius $R$. (The radius $R$ is proportional to $\sqrt{E}$
in the quantum problem.) One can easily produce a continuous
estimate ${\cal N}_W$ for such geometries \cite{Baltes76}. However,
the estimate of the {\em remainder} $r(E)$ dates back to ``Gauss circle problem"
(in 2D case), and has a long history of bounds \cite{Baltes76}
that are specific to the shapes of the quantum boxes.

Robinett studied a circular box in 2D confined in a sector~\cite{Robinett03}.
When the opening angle of the sector is changing, the area and the perimeter of
the confining box both change, but they are are not proportional to each other.
The structure of the eigenfunctions is relatively simple, since the azimuthal
(angular) eigenstates are simple sine functions. This work demonstrated the
validity of two-dimensional (2D) Weyl formula for the system. Particularly
enlightening in that study is the comparison of full (unconstricted) circle,
with a circle when the the sector has a angle of $2\pi$, i.e., degenerates
into a single excluding radius line, and with the sector that has opening
angle $\pi$, i.e., the particle is restricted to a semi-circular box. Our problem
of a spherical box restricted by a cone is the 3D generalization of the same
problem. However, as explained in the previous section, already the
determination of the polar degrees $\lambda$ as functions of the cone apex half-angle
$\theta_0$ had to be performed numerically, followed by a solution of the radial
eigenvalue equation determining the eigenenergies $E$, that are related to the
numerically known roots of Bessel functions. Nevertheless, we verified
the accuracy of Weyl expressions for several angles $\theta_0$. Unlike the
2D problem, the case corresponding to $\theta_0=\pi$, i.e., when the cone
becomes an excluded needle, the spectrum coincides with that of completely
unrestricted sphere, i.e., in 3D the excluded zero-width radial line does not
modify the energy spectrum. In this section we present in detail comparison
of the $\theta_0=\pi$ and $\theta_0=\pi/2$ cases, i.e., a complete sphere and
a hemisphere. The angular parts of the eigenfunctions in these cases are
represented by integer $\ell$s and $m$s and provide intuitive insights into
the properties of Weyl formula.

The fact that density of states {\em per unit volume} in 3D system exists
independently of the overall shape of the system, i.e., the leading term in
the number of states ${\cal N}$ is proportional to the system volume
$\cal V$, was implicit in the calculations of black body radiation or
counting mechanical oscillatory modes in a solid already in the late 19th
and early 20th centuries. Similar problem appears in the statistics of
the eigenenergies of Schr{\"o}dinger equation for a particle in a box.
Mathematically, this is a scalar Laplacian eigenvalue problem of determining
the number of eigenstates up to a certain value $E$. (It is related, but
not identical, to nonscalar problems, such as classical electromagnetic
or elastic waves, where $E$ is replaced by squared wavevector.)
It has been formally proven by Weyl that to the leading order $\cal N$
is proportional to the system volume $\cal V$. In our dimensionless
variables this number of states (for a scalar problem in 3D) is
${\cal T}_1=({\cal V}/6\pi^2)E^{3/2}$. Our choice of unit length scale $a$
does not affect the formula, because a choice of different length scale $a$,
modifies values of ${\cal V}$ and $E$, while making no change in the
coefficient of the formula. Similarly, the change in the {\em shape} of the
box does not influence this expression. In our examples of the sphere- and
hemisphere-shaped boxes the unit length defined as the radius of the
sphere, and the system volumes will be ${\cal V}_{\Circle}=4\pi/3$ and
$ {\cal V}_{\!\!\Rightcircle}=2\pi/3$, respectively.

For finite systems boundaries introduce subleading corrections to the total
number of states. The corrections depend on the type of boundary conditions
(b.c.) imposed on the wavefunction, such as function vanishing on the boundary
(Dirichlet b.c.), or normal derivative of the function vanishing on the boundary
(Neumann b.c.), or linear combination of the function and its normal derivative
vanishing (Robin b.c.) \cite{Balian70} conditions. In 1913 Weyl conjectured
\cite{Weyl13} that for smooth bounding surfaces the correction to the number of
eigenstates in the Laplacian problem with Dirichlet b.c. is proportional
to the {\em surface area} ${\cal S}$ and is given by ${\cal T}_2=-({\cal S}/16\pi)E$.
(The coefficient in this expression depends on b.c. \cite{Balian70} and, in
particular, for Neumann b.c. it is the same expression but with an opposite sign.)
In our examples of sphere- and hemisphere-shaped boxes the surface areas are
${\cal S}_{\Circle}=4\pi$ and $ {\cal S}_{\!\!\Rightcircle}=3\pi$, respectively.

Even smaller correction originates from the shape (``curvature") of the surface,
and is  given by ${\cal T}_3=({\cal C}/6\pi^2)E^{1/2}$. The {\em differentiable} parts
of the surface  where two main radii of curvature $R_1$ and $R_2$ can be defined
contribute to $\cal C$ the integral of  mean curvature
$\int dS\frac{1}{2}\left( 1/R_1+1/R_2\right)$. If the surface contains
sharp wedges, then their contribution depends on the wedge angle \cite{Baltes76},
and in particular  $90^\circ$ wedges contribute to ${\cal C}$ amount $(3\pi/8){\cal L}$,
where $\cal L$ is the total length of such wedges. In our examples
for a spherical box we have only the curvature term: since the mean
curvature of the sphere of unit radius is 1, the total curvature
contribution is ${\cal C}_{\Circle}=4\pi$. For the hemisphere
the nonvanishing curvature contributes $2\pi$, while the $90^\circ$
edge of length $2\pi$ contributes another $3\pi^2/4$, leading
to total ${\cal C}_{\!\!\Rightcircle}=2\pi+3\pi^2/4$.

\begin{figure}[t!]

 \begin{subfigure}[b]{0.45\textwidth}
 \center
        \includegraphics[width=8 truecm]{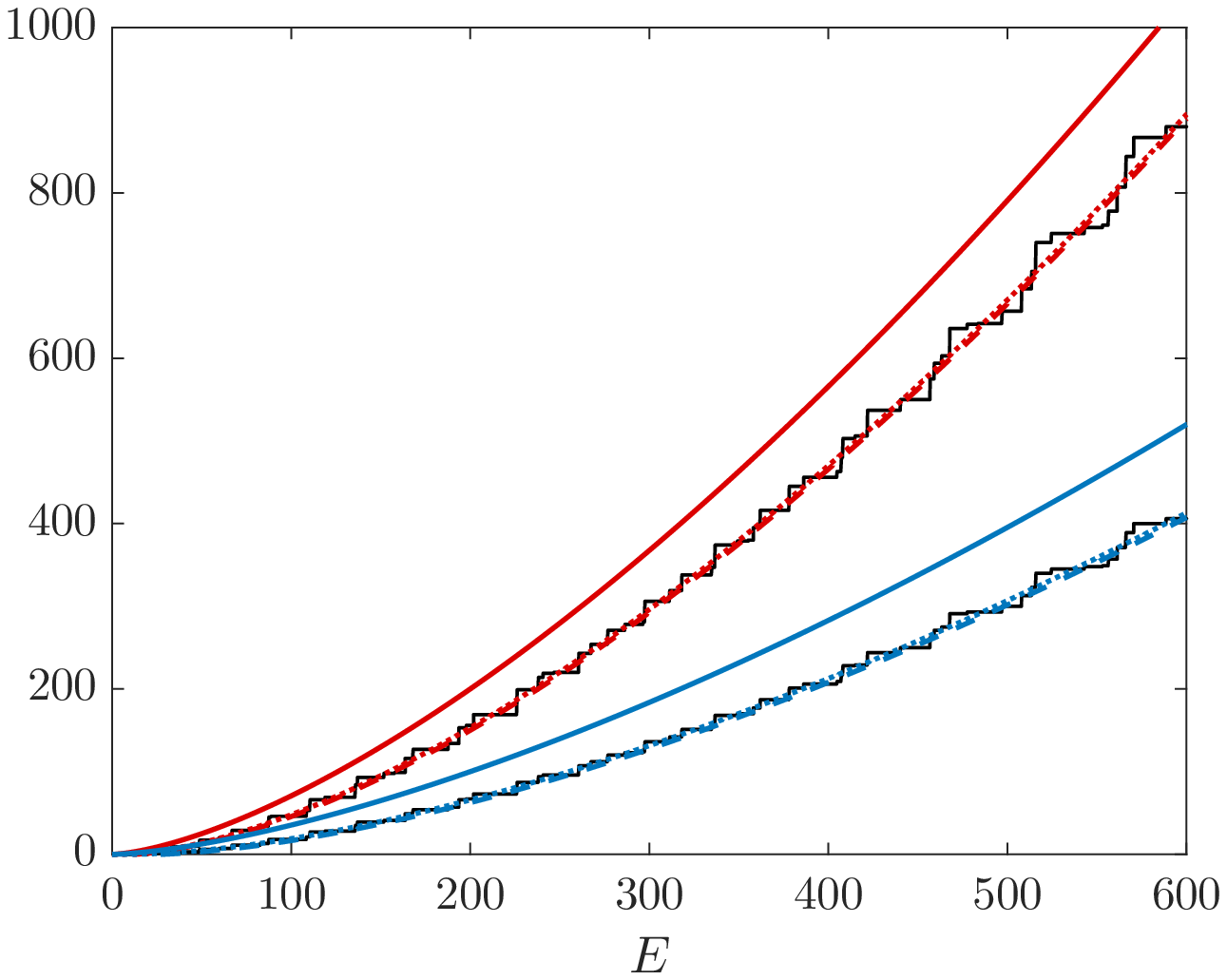}
      \caption{}
        \label{fig:N_vs_E}
    \end{subfigure}
    \hfill
     \begin{subfigure}[b]{0.45\textwidth}
     \center
        \includegraphics[width=8 truecm]{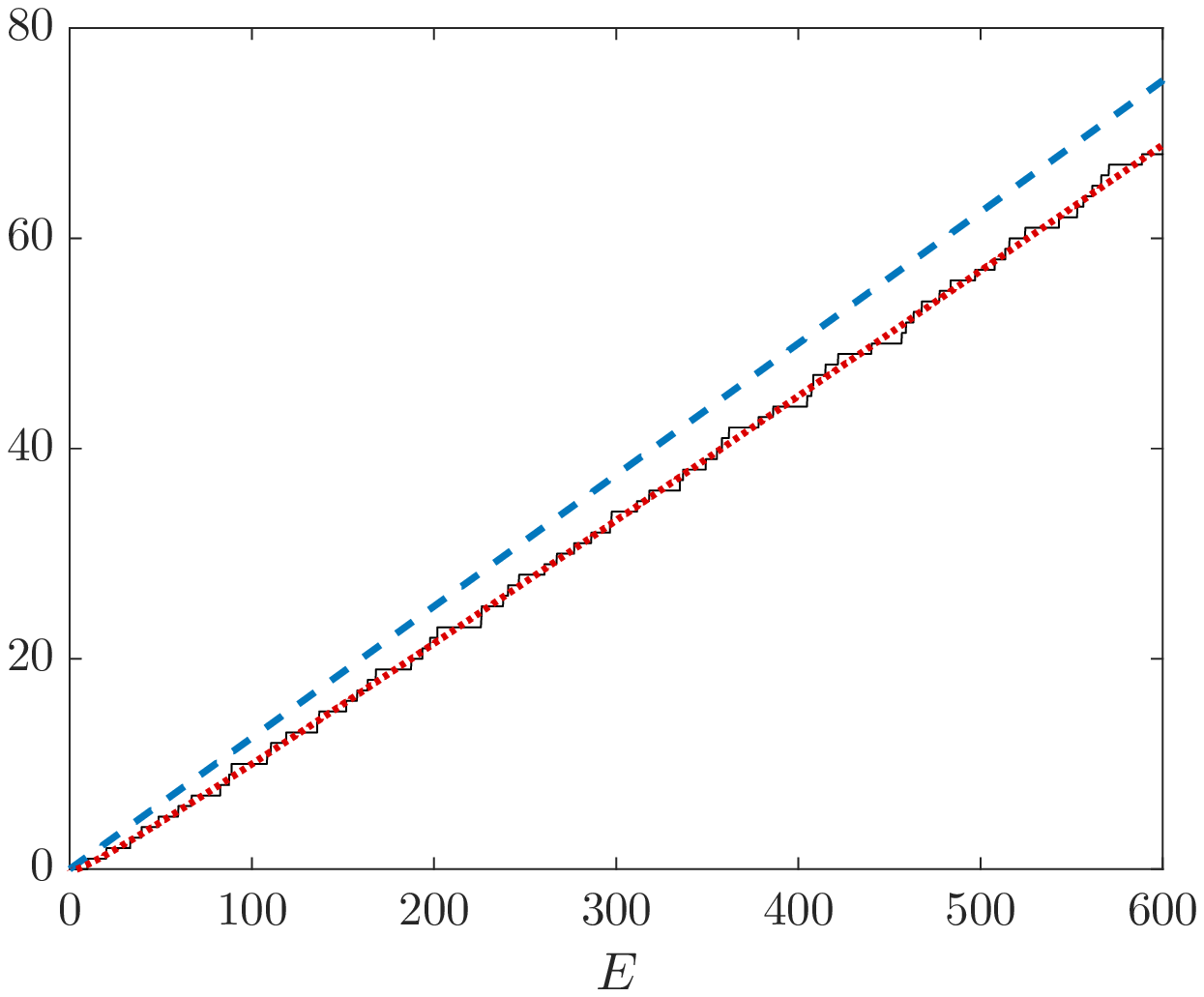}
       \caption{}
        \label{fig:N_difference}
    \end{subfigure}

\caption{(a) The exact numbers of eigenstates (step-wise solid black lines)
for sphere (top bundle of lines) and hemisphere (lower bundle of lines) as
functions of the energy $E$.
The exact results are compared with Weyl approximate continuous lines
for a sphere (red) and a hemisphere (blue) as in  Eqs.~\eqref{eq:N_sphere}
and \eqref{eq:N_hemisphere}, where in  each bundle the top solid lines
show only the first (leading) term for both geometries, the lowest
dashed lines show two first terms, and a  dotted lines slightly above
the dashed lines show all three terms. (b) The exact difference between
the numbers of steps in a sphere
and twice the number of states in a hemisphere (step-wise solid lines)
${\cal N}_{\Circle}-2{\cal N}_{\!\!\Rightcircle}$  as function of the energy $E$.
It is compared with the Weyl estimates of the difference. The leading terms in Weyl
formulas cancel out, and the dashed (blue) line represents the difference of
second terms ($\frac{1}{8}E$), while the dotted (red) line represents the difference
of second and third terms, i.e., $\frac{1}{8}E-\frac{1}{4}E^{1/2}$.
}
\end{figure}

By combining the volume, surface area and ``curvature" terms of Weyl
function we get the following expression for the number of
states~\cite{Baltes76}
\begin{equation}
{\cal N}_W(E)={\cal T}_1+{\cal T}_2+{\cal T}_3+\dots
=\frac{{\cal V}}{6\pi^2}E^{3/2}
-\frac{{\cal S}}{16\pi}E+\frac{{\cal C}}{6\pi^2}E^{1/2}+o(E^{1/2}),
\label{eq:Weyl}
\end{equation}
In the specific cases of sphere and hemisphere, these can be written as
\begin{align}
{\cal N}_{W\Circle}&=\frac{2}{9\pi}E^{3/2}
-\frac{1}{4}E+\frac{2}{3\pi}E^{1/2}+\dots
\label{eq:N_sphere}\\
{\cal N}_{W\!\!\Rightcircle}&=\frac{1}{9\pi}E^{3/2}
-\frac{3}{16}E+\left(\frac{1}{3\pi}+\frac{1}{8}\right)E^{1/2}+\dots
\label{eq:N_hemisphere}
\end{align}
The validity of these expressions is demonstrated in Fig.~\ref{fig:N_vs_E}
which has been obtained by numerically enumerating the total number of
states for each energy $E$. The figure presents the exactly measured
${\cal N}(E)$ both for the sphere and the hemisphere and compares the exact
results with Weyl functions in Eqs.~\eqref{eq:N_sphere} and
\eqref{eq:N_hemisphere}. Each of the latter equations are shown in three
approximate forms: solid line depicts only the first Weyl
terms, the dashed line depicts first two Weyl terms, and the dotted line
depicts three terms. While two-term lines represent strong improvement in
the correspondence with measured $\cal N$ over the single-term lines, the
three-term lines are barely distinguishable from two-term lines in
Fig.~\ref{fig:N_vs_E}.  Moreover, the fluctuations of the remainder $r(E)$
defined in Eq.~\eqref{eq:remainder} increase with $E$, and their departure
from the continuous curves is larger than the third term correction to
${\cal N}_W$. (A meaningful comparison and validation of the third term can
be done only if the exact stepwise curve $\cal N$ is smoothed by an averaging
procedure \cite{Baltes76}.) The extent of fluctuations can be characterized
by a bound $|r(E)|<cE^\beta$, where $c$ and $\beta$ are some constants.
For a cubic box there exist some theoretical bounds on $\beta$, but no
such bounds are known for a spherical box. By numerically examination the
$|r(E)|$ graphs in the range $E<4000$, we note that all data points of a sphere
fit under a curve with $\beta=1/2$, and $c\approx 3.2$, while for
the hemisphere we have the same $\beta$ with twice smaller $c$. (However, these
numbers are just the ``ballpark" estimates of a ``random" function in a limited
energy range.)

As explained in the previous section the eigenstates of a complete sphere and a
hemisphere can be represented by integer $l$s, and the energies become
independent of $m$.  Therefore both the exact $\cal N$s and their Weyl
approximations ${\cal N}_W$ are simply related. For a complete sphere each
$\ell$ state is $2\ell+1$ times degenerate, while in hemisphere the parity
of $\ell$ and $m$ must be opposite, and therefore each $\ell$ state is
degenerate $\ell$ times. Let ${\cal N}_\ell(E)$ be a number of zeroes of
the spherical Bessel function $j_l(x)$ up to certain maximal value $x=\sqrt{E}$,
which via relation \eqref{eq:E_relate_to_root} is the number of distinct
eigenenergies, up to energy $E$. (This function ignores the degeneracies
of the energies.) Then
\begin{align}
{\cal N}_{\Circle}(E)&=\sum_\ell (2\ell+1){\cal N}_\ell(E),
\label{eq:NsphereViaNell}\\
{\cal N}_{\!\!\Rightcircle}(E)&=\sum_\ell \ell{\cal N}_\ell(E).
\label{eq:NhemisphereViaNell}
\end{align}
(The summation over $\ell$ is finite since starting with some $\ell$ there
are no more eigenstates with energies lower than $E$.) It is known \cite{Arriola89}
that for large $\ell$s the density of zeros of spherical Bessel function $j_\ell$
approaches a constant: $\rho(x)=(1/\pi)(1+(\ell+1/2)^2/x^2)^{1/2}$ for $x>\ell+1/2$.
Consequently, for a given $\ell$ the number of roots up to energy $E$ will be
(to the leading order and neglecting prefactors) ${\cal N}_\ell(E)\sim(\sqrt{E}-\ell)$.
Both ${\cal N}_{\Circle}(E)$ and ${\cal N}_{\!\!\Rightcircle}(E)$ in
Eqs.~\eqref{eq:NsphereViaNell} and \eqref{eq:NhemisphereViaNell} include
$\sum_\ell \ell{\cal N}_\ell(E)\sim \int^{\sqrt{E}}d\ell\ell (\sqrt{E}-\ell)\sim E^{3/2}$,
thus explaining the leading terms in the Weyl relations.

It is interesting to consider the {\em difference} between the number of states
in a sphere and {\em twice} the number of states in a hemisphere. From
Eqs.~\eqref{eq:NsphereViaNell} and \eqref{eq:NhemisphereViaNell}
${\cal N}_{\Circle}(E)-2{\cal N}_{\!\!\Rightcircle}(E)=\sum_\ell{\cal N}_\ell(E)$.
Not surprisingly, this difference cancels out the terms proportional to
volume, and we can expect to have the leading term proportional to $E$. Indeed
from the approximate value of ${\cal N}_l$ we find that
$\sum_\ell {\cal N}_\ell(E)\sim \int^{\sqrt{E}}d\ell(\sqrt{E}-\ell)\sim E$.
Analogously, by subtracting Weyl functions of these systems we get
${\cal N}_{W{\Circle}}(E)-2{\cal N}_{W{\!\!\Rightcircle}}(E)=\frac{1}{8}E-\frac{1}{4}E^{1/2}$.
Figure \ref{fig:N_difference} depicts the exact difference between the numbers
of states by a step-wise solid line. This result is compared with
the difference obtained from Weyl expressions truncated at the second
term (dashed line) and including also the third term (dotted line).
We note an excellent agreement between the exact results and the predictions
of Weyl functions. Such clarity of the result is a consequence of very small
remainders $r(E)$ which are significantly smaller than were seen in
Fig.~\ref{fig:N_vs_E}. In fact in the range of $E<4000$ we found that
$|r(E)|<4$, i.e., {\em it does not increase with} $E$. Apparently,
since this is the sum of ${\cal N}_l$s that does not account for degeneracies
of the eigenstates, the fluctuations are significantly suppressed.
The accuracy of the third term in Weyl equation can be clearly seen
in this picture. It seems that the entire increase in the fluctuations $|r(E)|$
in Fig.~\ref{fig:N_vs_E} was caused by the growing degeneracy of the high
energy eigenstates.

\section{Finite spherical well bounded by a cone}\label{sec:FiniteWell}

It is well known in quantum mechanics that purely attractive potential in $d=1$
always has a bound state \cite{LL_vol3}, and for a sufficiently deep well
it may have many bound states. (There is also a slightly more relaxed
criterion guaranteeing the presence of bound states
\cite{Kocher77,Buell95,Brownstein00}.) Similar situation also exists in 2D,
where the bound state can always be found \cite{Chadan03}. If space dimension
$d$ is considered as continuous variable, it can be shown that this
property disappears when $d>2$ \cite{Nieto02}. In particular, in 3D
the presence or absence of the bound state depends on the shape and depth
of the binding potential. However, for potentials that have repulsive parts,
such as infinite barriers, the bound states are not necessarily present,
and 3D case may resemble lower-dimensional situations.

In this section we consider a \textit{finite} spherical well of radius $a=1$
with depth $U_0$ below zero potential outside the well,
measured in the same
dimensionless units as in
Sec.~\ref{sec:InfiniteWell}. The finite well is bounded by a circular cone
with an apex angle $2\theta_0$ such that
\begin{equation}
V\left(\text{\textbf{r}}\right)=\begin{cases}
-U_0, & r<1,\,\theta<\theta_{0},\\
0, & r>1,\,\theta<\theta_{0},\\
\infty, & \text{otherwise}.
\end{cases}
\end{equation}
This system admits both bound ($E<0$) and unbound ($E>0$) eigenstates, and
we will examine the transitions between them for various well depths and apex
half-angles $\theta_0$. Since the problem permits the same variable separation
as in the infinite well case discussed in Sec.~\ref{sec:InfiniteWell} the angular
(polar and azimuthal) dependence of both bound and unbound states inside
and outside the spherical well are determined by the cone and are identified
by the same $\lambda_i^m$s corresponding to a particular $\theta_0$, that
were depicted in Fig.~\ref{fig:lambda_vs_cos} and explained in detail in
Sec.~\ref{sec:InfiniteWell}.

The radial component of the wavefunctions $R(r)$ also satisfies the same
Eq.~\eqref{eq:radial} as for infinite potential well described in
Sec.~\ref{sec:InfiniteWell} but with the radial part of the potential
\begin{equation}
U(r)=
\begin{cases}
-U_0, & r<1,\\
0,  & r>1.
\end{cases}
\label{eq:FinitePotential}
\end{equation}
Within the regions that the potential is constant (either $-U_0$ or 0),
Eq.~\eqref{eq:radial} can be solved by {\em spherical} Bessel functions
of the first and second kind, $j_\lambda$ and $n_\lambda$, respectively,
when the eigenenergy exceeds the potential, and {\em modified spherical}
Bessel functions $i_\lambda$ and $\kappa_\lambda$ for eigenenegies below the
the potential. The spherical Bessel functions are related to the {\em regular
(nonspherical)} Bessel functions (denoted by the capital letters) by
$j_\lambda(x)=\sqrt{\pi/2x}J_{\lambda+1/2}(x)$,
$\kappa_\lambda(x)=\sqrt{\pi/2x}K_{\lambda+1/2}(x)$,
and the same for other function pairs. We note, that the regular Bessel functions
$J_\lambda$, $N_\lambda$, $I_\lambda$, and $K_\lambda$ solve analogous
radial equation in 2D \cite{HLKK_PRE96}, and some of the results presented
below resemble solutions of 2D problem, with $\lambda$ shifted by $1/2$.

For bound states ($-U_0<E<0$) the radial part of the wavefunction is proportional
to $j_\lambda(kr)$, with $k=\sqrt{E+U_0}$ inside the well, while $n_\lambda$ is
dismissed since it diverges at the  origin, and $\kappa_\lambda(qr)$, with
$q=\sqrt{-E}$ outside the well, while $i_\lambda$ is dismissed since it diverges
at $r\to\infty$. Thus the wavefunction has the form
\begin{equation}
\psi_{n\lambda m}({\bf r})\propto
P_{\lambda}^{m}\left(\cos\theta\right)e^{im\varphi}
\begin{cases}  j_\lambda\left(kr\right), & r<1,\\
          \kappa_\lambda\left(qr\right), & r>1.
\end{cases}
 \end{equation}
Since the radial part of the Schr{\"o}dinger equation is second order
differential equation with a potential which is a stepfunction ar $r=1$,
both the wave function and its derivative must by continuous at $r=1$,
although the slope of the derivative changes at $r=1$ leading to a finite
jump in the second derivative, and therefore
\begin{equation}
\frac{kj'_\lambda(k)}{j_\lambda(k)}=\frac{qk'_\lambda(q)}{\kappa_\lambda(q)}\ ,
\label{eq:continuity}
\end{equation}
where prime denotes derivative of the function. The overall prefactor
of the functions is determined from normalization conditions. Since
the value of $\lambda=\lambda_i^m$ was determined form the angular equation, and
$U_0$ is implicit in the definition of $k$, the only unknown parameter
in Eq.~\eqref{eq:continuity} is energy $E$ that determines both $k$ and $q$.
The possible values of $E$ satisfying this equation are the eigenenegies of
the bound eigenstates. Interestingly enough, if we express the spherical Bessel
functions via regular ones and perform the derivatives in the numerators
of all the functions, Eq.~\eqref{eq:continuity} becomes the relation
\begin{equation}
\frac{kJ'_{\lambda+1/2}(k)}{J_{\lambda+1/2}(k)}
=\frac{qK'_{\lambda+1/2}(q)}{K_{\lambda+1/2}(q)} ,
\label{eq:continuity_regular}
\end{equation}
which is exactly the 2D continuity relation for a circular well contained by a
sector, but with a shift $\lambda\to\lambda+1/2$ \cite{HLKK_PRE96}.

\begin{figure}[t!]

 \begin{subfigure}[b]{0.45\textwidth}
 \center
        \includegraphics[width=8 truecm]{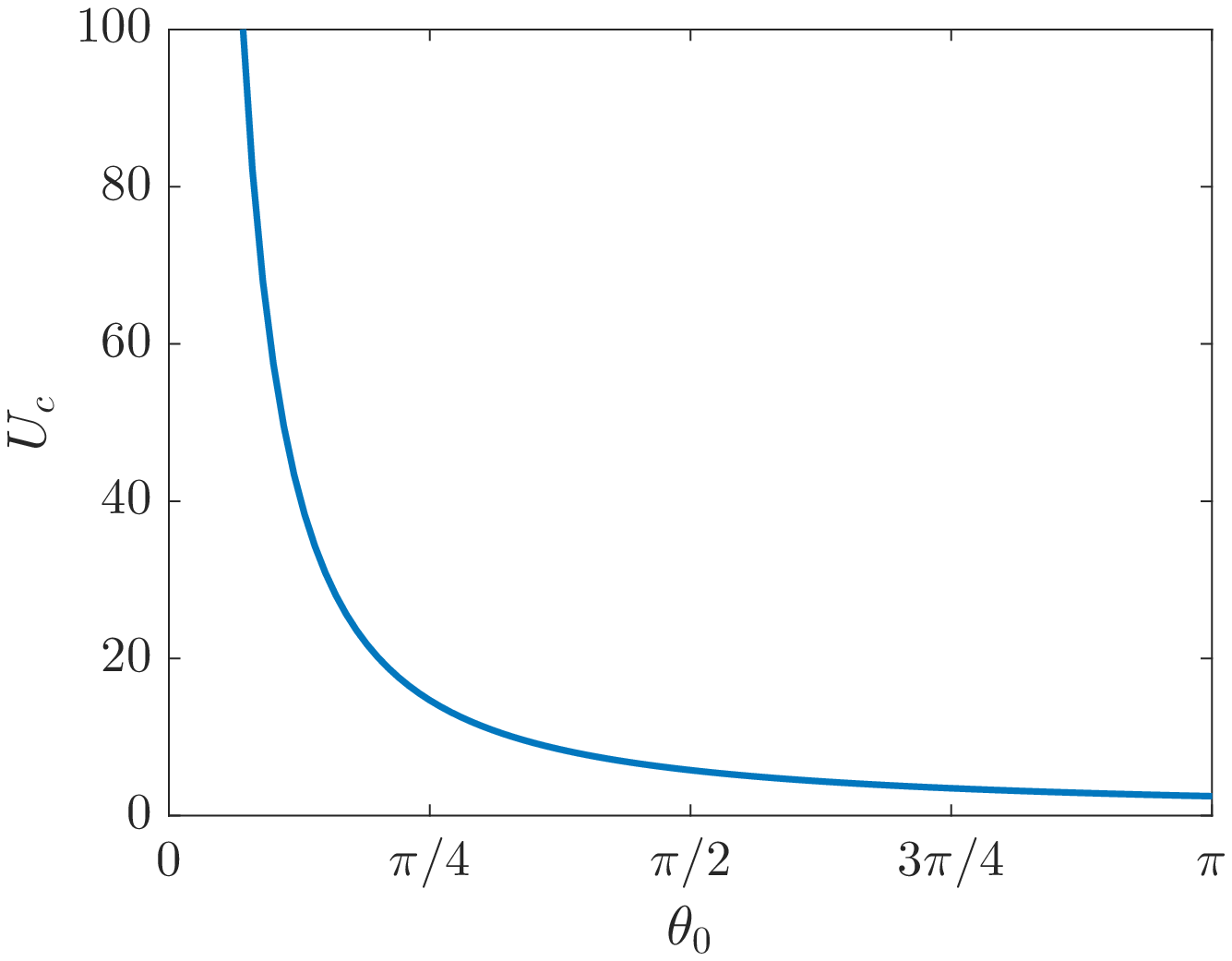}
        \caption[Network2]%
        {}
        \label{fig:U_c_vs_theta_a}
    \end{subfigure}
    \hfill
     \begin{subfigure}[b]{0.45\textwidth}
     \center
        \includegraphics[width=8 truecm]{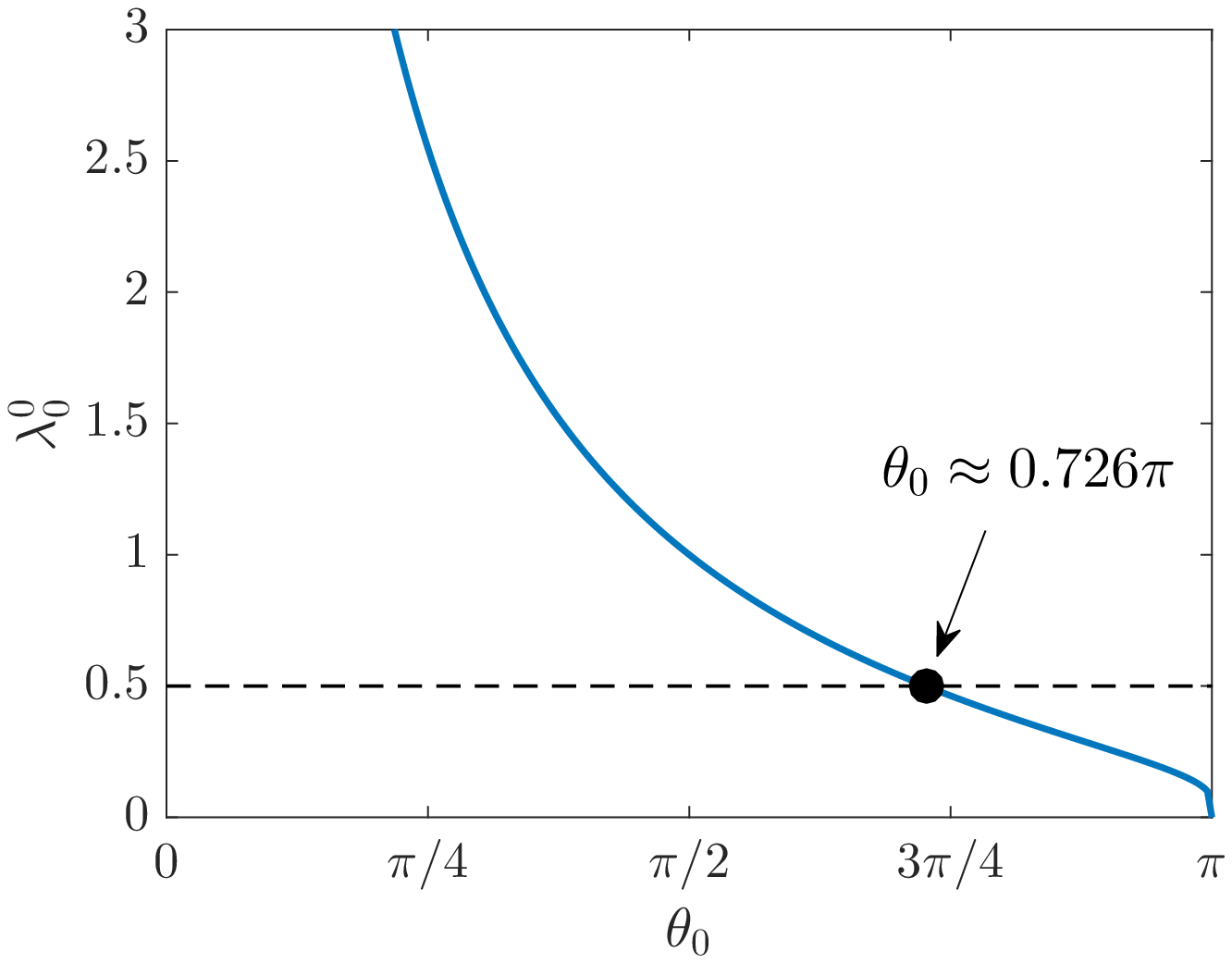}
        \caption[Network2]%
        {}
        \label{fig:lambda_vs_theta_b}
    \end{subfigure}

\caption{(a) The critical depth of the well $U_c=\alpha_1^2(\lambda_0^0-1)$
(see text) as a  function of the cone apex half-angle $\theta_0$. (b) The
ground-state angular degree $\lambda_0^0$ as a function of the cone apex
half-angle $\theta_0$. The special value  $\lambda_0^0=1/2$ is shown by a
dashed line.
}\label{fig:U_c_vs_theta}
\end{figure}

For various $\lambda_i^m$s there can be several, one or no bound eigenenergy
solutions of Eqs.~\eqref{eq:continuity} or \eqref{eq:continuity_regular}. As
the well becomes shallower ($U_0$ decreases) the number of bound states also
decreases, until only one bound eigenstate corresponding to $\lambda=\lambda_0^0$
remains with some eigenenergy $E_0<0$. When the well depth decreases to some
critical $U_c$ the bound states disappear altogether. When $E_0\to0$, then $q\to0$,
and the right hand side of Eq.~\eqref{eq:continuity_regular} approaches
$-(\lambda+1/2)$ \cite{HLKK_PRE96}, while in the left hand side $k\to \sqrt{U_c}$.
The relation for critical depth of the well is
\begin{equation}
\frac{\sqrt{U_c}J'_{\lambda+1/2}(\sqrt{U_c})}{J_{\lambda+1/2}(\sqrt{U_c})}=-(\lambda+1/2),
\label{eq:U_c}
\end{equation}
which by using recurrence relations between Bessel functions and their derivatives
\cite{Abramowitz72} reduces to
\begin{equation}
J_{\lambda-1/2}(k)=0, \ {\rm or}\ \ j_{\lambda-1}(k)=0,
\label{eq:zero}
\end{equation}
with $k=\sqrt{U_c}$. Thus $U_c$ is simply the square of the first zero of these
functions. As can be seen in Fig.~\ref{fig:lambda_vs_cos} for each $\theta_0$
(or $w=\cos\theta_0$) there exists an infinite sequence of $\lambda_i^m$s.
Since, $U_c$ represents the case when the last remaining (ground) eugenstate
has zero energy, we must use the lowest $\lambda=\lambda_0^0$, and therefore
\begin{equation}
U_c=\alpha^2_1(\lambda_0^0-1).
\label{eq:U_c2}
\end{equation}
(To compare this result with the 2D case, see Eq.~(16) in Ref.~\cite{HLKK_PRE96}.)
Fig.~\ref{fig:U_c_vs_theta}a depicts the dependence of the critical depth $U_c$
on the cone apex half-angle $\theta_0$. For small $\theta_0$ the confinement is
strong and large $U_c$ is required. Indeed, for large $\lambda$ the first root of
$J_\lambda$ is approximately $\lambda$, and therefore $U_c\sim\lambda^2\sim1/\theta_0^2$.
As $\theta\to\pi$ the critical depth drops to the critical value of an unconstrained
spherical well.

Equations  \eqref{eq:U_c} and \eqref{eq:zero} relied solely on the assumption that
the energy of the eigenstate vanishes and were not specific to the case of single
remaining bound state. We might consider situations when a vanishing energy eigenstate
appears for deeper wells, when additional negative energy bound states are
still present. For each $\lambda_i^m$ there is an infinite amount of such
well depths corresponding to different zeros of the same Bessel function:
\begin{equation}
U_0^{(i,m,n)}=\alpha^2_n(\lambda_i^m-1).
\label{eq:U_cim}
\end{equation}
The critical $U_c$ in Eq.~\eqref{eq:U_c2} is just the smallest depth in the
infinite sequence of values in Eq.~\eqref{eq:U_cim}.

\section{Critical exponents}\label{sec:Exponents}

The radial part of the wavefunction of a bound state is characterized
by energy $E<0$ and the corresponding $q=\sqrt{-E}$ outside the well
is described by $\kappa_\lambda(qr)$, where the polar constant (degree)
$\lambda=\lambda_i^m$, while the energy (or $q$) is obtained from one
of the solutions of Eq.~\eqref{eq:continuity}. For $qr\gg1$, the function
$\kappa_\lambda(qr)\sim \frac{1}{qr} e^{-qr}$, i.e., it has a typical decay
(localization) length $\xi\equiv 1/q=1/\sqrt{-E}$. As the energy of a specific
eigenstate approaches 0 the length $\xi\to\infty$. Eigenstates with zero
binding energy for several potential types have been studied in
Refs.~\cite{Daboul94,Hojman95}, and became parts  of textbooks (see, e.g.,
Ref.~\cite{Schiff68}). For the $E=0$  eigenstate the radial part of the
eigenfunction becomes
\begin{equation}
R(r)\propto
\begin{cases}
j_{\lambda}\left(kr\right), & r<1,\\
1/r^{\lambda+1}, & r>1,
\end{cases}
\end{equation}
where the continuity condition at $r=1$ in Eq.~\eqref{eq:continuity} is
the same as in Eqs.~(\ref{eq:continuity}-\ref{eq:zero}). If we are
considering the single surviving bound state, then $\lambda=\lambda_0^0$,
$k=\sqrt{U_c}$, which is given  by the critical value $U_c$ defined in
Eq.~\eqref{eq:U_c2}. This case plays an especially important role, because in
the analogy between quantum states and ideal polymers, the state of long
polymer is dominated by the {\em ground state} of the quantum problem, and whether
this state is bound or not determines whether the polymer is adsorbed or not to
the attractive potential area. (See Sec.~\ref{sec:Discussion}.)
However, the argument presented here can also
be applied for deeper potentials $U_0$ that support several bound states,
for each case of the depth $U_0=U_0^{(i,m,n)}$ as defined in Eq.~\eqref{eq:U_cim}
for which a zero energy state appears. For simplicity, below we will
mostly consider only a well slightly deeper than $U_c$ and admitting a single
bound state.

\begin{figure*}[t!]
\center
\includegraphics[width=15 truecm]{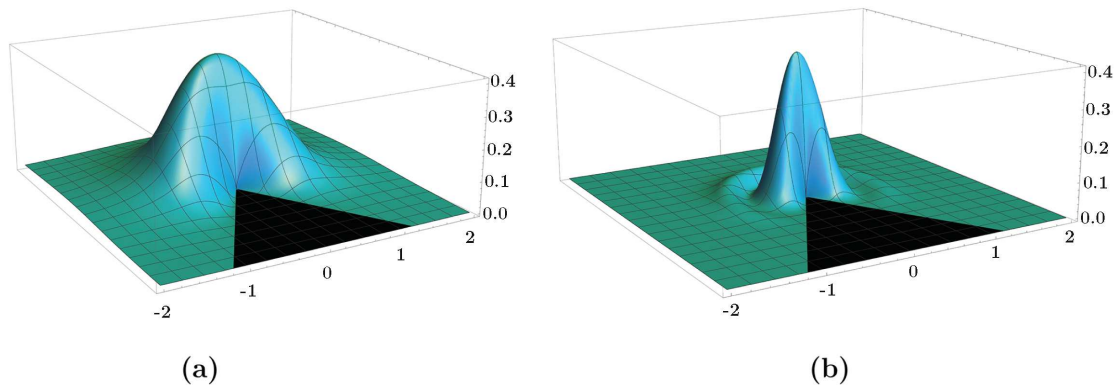}
\caption{Probability density function $|\psi (r,\theta,\varphi=0)|^2$ (not normalized)
for $\theta_0=5\pi/6\approx 0.833\pi$. The horizontal plane of the graph represents
$\theta,r$ coordinate plane, i.e., a cut through $x-z$ plane of Cartesian coordinates.
The excluded cone is represented by the darker area of the plane. Two cases with two
different eigenenergies are shown: (a) $E=0$ eigenstate for $U_0=U_c\approx4.740$,
corresponding to $\lambda_0^0\approx0.355$ with $\psi$ that decays as a power law
$\sim 1/r^{1.355}$ outside the sphere, and therefore is not square-integrable.
(b) Bound eigenstate with $E\approx-7.613$ corresponding to the second zero of the Bessel function
for $U_0=40\approx 8.44U_c$ and the same $\lambda_0^0$, that decays exponentially for
large distances. While this is the second state of $\lambda_0^0$, it is the sixth lowest state
overall.
}\label{fig:psi2}
\end{figure*}

The normalization of the ground state requires the radial integration
$\int r^2 |\psi|^2 dr$. If the radial part $R(r)$ of the eigenstate
behaves at large distances as $1/r^{\lambda+1}$,  the integral will
only converge for $\lambda>1/2$. Figure \ref{fig:lambda_vs_theta_b}
redraws the lowest branch $\lambda_0^0$ of Fig.~\ref{fig:lambda_vs_cos}
as a function of $\theta_0$.  There is an entire region of this branch
where $0<\lambda_0^0\le 1/2$ corresponding
$\pi>\theta_0>\theta_c\approx0.726\pi$. In that region the normalization
cannot be performed, and these $E=0$ energy states can be categorized
as unbound states.  Figure \ref{fig:psi2}a depicts such a
{\em non-normalizable} state for $U_0=U_c$ at large $\theta_0$.
Interestingly enough, $E=0$ states corresponding to other branches of
$\lambda$ spectrum always have $\lambda_i^m\ge1$ and therefore are
normalizable. For deep $U_0$ there may be other eigenstates
corresponding to $\lambda_0^0$ and higher order zeros of of Bessel
function, which also cannot be normalized at these angles. The bound
state depicted in Fig.~\ref{fig:psi2}b corresponds (for large $U_0$
to the {\em second} zero of Bessel function for the same $\lambda_0^0$
is the eigenstate in Fig.~\ref{fig:psi2}a. The fact that it is the
second zero can be seen in the single oscillation that performs the
wavefunction inside the well. If $U_0$ is decreased, at some point the
eigenenergy of this state will reach zero and it will resemble the
behavior of the state in Fig.~\ref{fig:psi2}a but with larger $k$ in
the $r<1$ region. Similar type of normalizability problem exists in 2D
problem of a sector confining a circle. However, in 2D the borderline
$\theta_c$ corresponds to the sector becoming a straight line, while
in our case the special angle of $\theta_c$ has no ``special"
geometrical meaning.

For $U_0>U_c$ the ground state has finite $q$ and $\xi$ which we expect to vanish and
diverge, respectively, as $U_0\to U_c$. When $\delta U\equiv U_0-U_c\ll U_c$, then $E$
is also small and we can expand both sides of Eq.~\eqref{eq:continuity} in these small
quantities to find the dependence of $E$ on $\delta U$, and therefore the dependence
of $\xi$ on $\delta U$. We denote the left hand side of Eq.~\eqref{eq:continuity}
by ${\cal G}_{\text{L},\lambda}(k)\equiv kj'_{\lambda}(k)/j_{\lambda}(k)$,
and right hand side by
${\cal G}_{\text{R},\lambda}(q)\equiv q\kappa'_{\lambda}(q/\kappa_{\lambda}(q)$.
The expansion of the left hand side is given by \cite{Abramowitz72}
\begin{equation}
{\cal G}_{\text{L},\lambda}(k)={\cal G}_{\text{L},\lambda}\left(\sqrt{U_c+\delta U+E}\right)
\approx-\lambda-1-\frac{1}{2}(E+\delta U),
\end{equation}
while the form of the expansion of the right hand side depends on the value of
$\lambda$ \cite{Abramowitz72}:
\begin{equation}
{\cal G}_{\text{R},\lambda}(q)={\cal G}_{\text{R},\lambda}\left(\sqrt{-E}\right)
\approx\begin{cases}
-\lambda-1-\frac{1}{2\left(\lambda-\frac{1}{2}\right)}\left(-E\right), & \lambda>\frac{1}{2},\\
-\frac{3}{2}+\frac{1}{2}\left(-E\right)\ln\left(-E\right), & \lambda=\frac{1}{2},\\
-\lambda-1-\frac{2^{-2\lambda}\pi}{\cos\left(\pi\lambda\right)
\Gamma^{2}\left(\lambda+\frac{1}{2}\right)}\left(-E\right)^{\lambda
+\frac{1}{2}}, & 0\leq\lambda<\frac{1}{2}.
\end{cases}
\end{equation}
By equating ${\cal G}_{\text{L},\lambda}={\cal G}_{\text{R},\lambda}$ we extract
the dependence of $-E$ on $\delta U$, and consequently the $\delta U$-dependence of $\xi$,
in the form $\xi\sim\delta U^{-\nu}$:
\begin{equation}
\xi\sim\begin{cases}
\delta U^{-\frac{1}{2}}, & \text{for}\,\lambda>\frac{1}{2},\\
\left|\ln\delta U\right|^{\frac{1}{2}}\delta U^{-\frac{1}{2}}, & \text{for}\,\lambda=\frac{1}{2},\\
\delta U^{-\frac{1}{2\lambda+1}}, & \text{for}\,0\leq\lambda<\frac{1}{2}.
\end{cases}
\label{eq:xi_vs_delta_U}
\end{equation}
For small angles $\theta_0$ (large $\lambda$) the critical exponent controlling the
$\delta U$-dependence of $\xi$ is $\nu=1/2$. However for large enough angles
$\theta_0>\theta_c$ the exponent $\nu=1/(2\lambda+1)$ becomes angle $\theta_0$-dependent
and reaches 1 when $\theta_0\to\pi$. Note that the transition between angle-dependent
and angle-independent regimes occurs when at $\theta_0=\theta_c$ when $\lambda=1/2$.

The above derivation could be repeated also in the situation when $U_0$ approaches
any of the $U_0^{(i,m,n)}$ defined in Eq.~\eqref{eq:U_cim}, and the
Eq.~\eqref{eq:xi_vs_delta_U} is valid with $\delta U\equiv U_0-U_0^{(i,m,n)}$.
However, the only $\lambda$ capable of having values below $1/2$ is $\lambda_0^0$.
Therefore, the usual value of the exponent is $\nu=1/2$, with exception
of the cases of $\theta_0>\theta_c$ and $\lambda_0^0<1/2$, corresponding to various
order zeros, i.e., $U_0^{(0,0,n)}$ with $n=1,2,\dots$

\section{Discussion}\label{sec:Discussion}

Some of the results derived in our work resemble the regular solutions
of a particle in a spherical box or a particle in a finite potential
well in the absence of the cone. However, in the spherically-symmetric
case the degree $\ell$ is integer, and many of the effects described
in this work are absent. For instance, the discussion of $E=0$
eigenstates in Refs.~\cite{Daboul94,Hojman95} omits the trivial
$\ell=0$ case, and proceeds to discuss $\ell\ge1$ case, when the
interesting effects and the qualitative change in the behavior, such as
lack of integrability of the $E=0$ state, appears for noninteger
$\lambda\le 1/2$.

In the mapping between the quantum and the ideal polymer
problems~\cite{Gennes69} the quantum potential is replaced by the actual
potential divided by $k_BT$. The probability density of the
end-point of an $N$-monomer polymer is given by a superposition of the
${\bf r}$-dependent quantum eigenstates multiplied by ${\rm e}^{-EN}$,
where $E$ is energy of a particular state and the exponent replaces the
time-depended oscillatory term of the quantum mechanics. For large  $N$ the
state with the smallest $E$ will dominate the density distribution and its
localization length $\xi$ will be the spatial extent of the polymer.
Since $\xi$ is finite only for the bound quantum states, the absence or
presence of the polymer adsorption is related to the existence of
the bound state. The
effective depth of the ``polymer potential" is varied by changing $T$,
and when $T$ approaches the critical adsorption temperature $T_a$ the
effective value $U$ of the quantum problem approaches $U_c$
and $\delta U\sim (T_a-T)$. In this situation
Eq.~\eqref{eq:xi_vs_delta_U} can be re-written in the form
\begin{equation}
\xi\sim\begin{cases}
(T_a-T)^{-\frac{1}{2}}, & \text{for}\,\lambda>\frac{1}{2},\\
\left|\ln(T_a-T)\right|^{\frac{1}{2}}(T_a-T)^{-\frac{1}{2}}, & \text{for}\,\lambda=\frac{1}{2},\\
(T_a-T)^{-\frac{1}{2\lambda+1}}, & \text{for}\,0\leq\lambda<\frac{1}{2},
\end{cases}
\label{eq:xi_vs_delta_T}
\end{equation}
which makes it an interesting {\em thermal} phase transition for
{\em very long} polymers, with possibly cone apex angle-dependent critical exponent.
Since the behavior of the ideal polymer is dominated by the ground state,
Eq.~\eqref{eq:xi_vs_delta_T} can be used only for $\lambda=\lambda_0^0$
and only in the neighborhood of $U_c$, which is related to the {\em first}
zero of the Bessel function as in Eq.~\eqref{eq:U_c2}. The behavior
of real polymers in good solvents, where the monomers strongly repel
each other, follows the behavior of the ideal polymers only
qualitatively. Even when the quality of the solvent is decreased
and effectively cancels out the monomer repulsion (in so-called
``$\theta$-solvents" \cite{degennesSC}), the correspondence to
ideal polymers is only approximate. Moreover, for a very long
polymer the adsorption in a finite-volume spherical well is geometrically
impossible. However, the polymers of moderate stiffness (such as DNA)
have a broad range of length-scales where the rigidity can be neglected,
while the mutual repulsion of the monomers is still very weak, that emulates long
ideal polymers~\cite{Nepal13}, and therefore can exhibit the transition effects.

\section*{Acknowledgments}
This work was supported by the Israel Science Foundation Grant No.~453/17.

\end{document}